\documentclass[preprint,number,sort&compress,5p,twocolumn,times]{elsarticle}

\usepackage[english]{babel}
\usepackage{textcomp}

\usepackage{graphicx}
\usepackage{xcolor}

\usepackage{amssymb}
\usepackage{amsmath}

\usepackage{lineno}

\usepackage{hyperref}
\usepackage{cleveref}

\biboptions{square}

\newcommand{\MeVee}{\ensuremath{\,\text{MeV}_\text{ee}}}
\newcommand{\MeV}{\ensuremath{\,\text{MeV}}}
\newcommand{\MHz}{\ensuremath{\,\text{MHz}}}
\newcommand{\Hz}{\ensuremath{\,\text{Hz}}}
\newcommand{\vect}[1]{\vec{#1}}

\journal{Nuclear Instruments and Methods in Physics Research A}

\newcommand*\patchAmsMathEnvironmentForLineno[1]{%
  \expandafter\let\csname old#1\expandafter\endcsname\csname #1\endcsname
  \expandafter\let\csname oldend#1\expandafter\endcsname\csname end#1\endcsname
  \renewenvironment{#1}%
     {\linenomath\csname old#1\endcsname}%
     {\csname oldend#1\endcsname\endlinenomath}}%
\newcommand*\patchBothAmsMathEnvironmentsForLineno[1]{%
  \patchAmsMathEnvironmentForLineno{#1}%
  \patchAmsMathEnvironmentForLineno{#1*}}%
\AtBeginDocument{%
\patchBothAmsMathEnvironmentsForLineno{equation}%
\patchBothAmsMathEnvironmentsForLineno{align}%
\patchBothAmsMathEnvironmentsForLineno{flalign}%
\patchBothAmsMathEnvironmentsForLineno{alignat}%
\patchBothAmsMathEnvironmentsForLineno{gather}%
\patchBothAmsMathEnvironmentsForLineno{multline}%
}

\addto\captionsenglish{
  
}

\begin{document}

\begin{frontmatter}



\title{Particle identification using clustering algorithms}


\author{R.~Wirth\corref{cor1}}
\ead{R.Wirth@gsi.de}

\author{E.~Fiori\corref{}}
\author{B.~L\"oher}
\author{D.~Savran}
\author{J.~Silva}

\address{ExtreMe Matter Institute EMMI and Research Division, GSI Helmholtzzentrum f\"ur Schwerionenforschung, Planckstr. 1, 64291 Darmstadt, Germany}
\address{Frankfurt Institute for Advanced Studies FIAS, Ruth-Moufang-Str. 1, 60438 Frankfurt am Main, Germany}

\author{H.~\'Alvarez Pol}
\author{D.~Cortina Gil}
\author{B.~Pietras}

\address{Departamento de F\'isica de Part\'iculas, Facultad de F\'isica, Universidade de Santiago de Compostela, 15782 Santiago de Compostela, Spain}

\author{T.~Bloch}
\author{T.~Kr\"oll}

\address{Institut f\"ur Kernphysik, Fachbereich 5 -- Physik, TU Darmstadt, Schlossgartenstr. 9, 64289 Darmstadt, Germany}

\author{E.~N\'acher}
\author{\'A.~Perea}
\author{O.~Tengblad}

\address{Instituto de Estructura de la Materia CSIC, Madrid, Spain}

\author{M.~Bendel}
\author{M.~Dierigl}
\author{R.~Gernh\"auser}
\author{T.~Le~Bleis}
\author{M.~Winkel}

\address{Physik-Department E12, TU M\"unchen, James-Franck-Stra{\ss}e, 85748 Garching, Germany}

\cortext[cor1]{Corresponding address: ExtreMe Matter Institute EMMI and Research Division, GSI Helmholtzzentrum f\"ur Schwerionenforschung, Planckstr. 1, 64291 Darmstadt, Germany}

\begin{abstract}
A method that uses fuzzy clustering algorithms to achieve particle identification based on pulse shape analysis is presented.
The fuzzy c-means clustering algorithm is used to compute mean (principal) pulse shapes induced by different particle species in an automatic and unsupervised fashion from a mixed set of data.
A discrimination amplitude is proposed using these principal pulse shapes to identify the originating particle species of a detector pulse.
Since this method does not make any assumptions about the specific features of the pulse shapes, it is very generic and suitable for multiple types of detectors.
The method is applied to discriminate between photon- and proton-induced signals in CsI(Tl) scintillator detectors and the results are compared to the well-known integration method.
\end{abstract}


\end{frontmatter}


\section{Introduction}
\label{sec:intro}
In many experimental setups, particle identification (PID) is an important task.
A well-known approach to identify the species of a particle hitting a detector is the use of pulse shape analysis (PSA), which exploits that the detector response depends on the particle species.

A very widely-used method for achieving particle identification using PSA is the integration method (see e.g.\ \cite{Alexander1961,Adams1978,Kaschuck2005}), which uses the ratio of the integrals over two regions of the detector signal to distinguish between different particle species.
A property of this method is that the regions that provide the best distinction have to be determined beforehand and---in order to achieve the best results---separately for each individual detector.
As detectors age and are exposed to radiation, their responses might change over time.
Changes of external parameters (e.g.\ temperature) might also affect the pulse shape properties and require a recalibration of parameters.
Thus, this calibration procedure, which requires human supervision, has to be repeated regularly.

For very large setups with thousands of detectors, e.g.\ the CALIFA barrel \cite{CalifaTDR} of the R\textsuperscript3B experiment at FAIR\footnote{Facility for Antiproton and Ion Research}, this is hardly feasible.
This is also true for detectors used in environmental field inspection with strongly varying conditions.
For this reason, it is highly desirable to develop a procedure that requires the least human supervision possible to perform the PID calibration of a detector system.

We have recently shown that fuzzy clustering algorithms can be used to perform the calibration for $\gamma$-neutron discrimination in a fully automatic and unsupervised manner for liquid scintillator detectors \cite{Savran2010}.
Due to the success of this method, we apply the same prescription in this work to the $\gamma$-proton discrimination using scintillating CsI(Tl) crystals, which have been shown to have distinct scintillation characteristics for different incident particles \cite{Alderighi2002,Benrachi1989,Skulski2001}.
We furthermore derive a means of using the information obtained from the clustering to perform the PID which is suitable for use in a signal processing environment, e.g.\ for online-identification using fast digital electronics.
In addition, the proposed procedure is independent of the particular features of the pulse shapes and thus represents a general approach suitable not only for CsI(Tl) detectors, but also for other detector types.

This paper is structured as follows:
\cref{sec:intro:fuzzy} gives a short introduction on the topic of clustering algorithms and outlines the algorithm on which this manuscript concentrates.
In \cref{sec:experiment}, the experimental setup is described during which the set of signals was recorded that we use to demonstrate the method.
\Cref{sec:preparation} focuses on the preprocessing of the raw signals that is employed in order to get stable and correct clustering results.
The results obtained from the clustering process are then used in the prescription described in \cref{sec:pid}.
There, we also summarize the well-known integration method and present a different approach, based on a specific distance measure, that does not need the results from the clustering process.

\subsection{The fuzzy c-means algorithm}
\label{sec:intro:fuzzy}
Clustering algorithms provide a means of assigning a set of $N$ vectors in $L$-dimensional space to $C$ groups, called clusters.
This assignment is done based on a notion of the distance between a vector from the set and the centroid of a cluster.
The cluster centroids themselves are not known beforehand, but are determined by the algorithm.

The degree of membership of the $i$-th vector to the $j$-th cluster is designated by a membership value $u_{ij}$, which is normalized so that the total degree of membership $\sum_{j=1}^C u_{ij}=1$ for every vector.
Based on the range of the membership values, clustering algorithms may be distinguished into two categories:
\begin{itemize}
 \item boolean clustering algorithms with $u_{ij}\in \{0,1\}$, i.e. a vector belongs exactly to one cluster.
 \item fuzzy clustering algorithms that use the full range of $u_{ij}\in [0,1]$ and thus can assign a single vector to multiple clusters with different degrees of membership.
\end{itemize}
In the following we will focus on fuzzy clustering algorithms because they comprise the more general case.

The concrete fuzzy clustering algorithm we successfully employed earlier \cite{Savran2010} is the fuzzy c\nobreakdash-means algorithm \cite{Bezdek1981,Duda2000}.
It can be posed as a minimization problem
\begin{equation}
 \label{eq:fcm-opt}
 \begin{aligned}
  \min_{u_{ij},\vect{c}_j}\, &J_m=\sum_{i=1}^{N} \sum_{j=1}^{C} u_{ij}^m \|\vect{x}_i - \vect{c}_j\|^2 \\
  \text{subject to} & \\
  &\sum_{i=1}^{N} u_{ij} > 0,\; j\in\{1,\dotsc,C\}\\
  &\sum_{j=1}^{C} u_{ij} = 1,\; i\in\{1,\dotsc,N\}\;,
 \end{aligned}
\end{equation}
i.e. minimizing the total intra-cluster distance $J_m$ subject to the constraints that no cluster should be empty and that the total degree of membership of any vector should be equal to unity.
Here, $\vect{x}_i$ denotes the $i$-th vector of the set, $\vect{c}_j$ is the centroid of the $j$-th cluster and $\|\cdot\|$ is the Euclidean norm; $N$ and $C$ denote the number of vectors and clusters, respectively.
Note that $C$ is a parameter of the algorithm that has to be known beforehand.

The parameter $m\in [1,\infty)$ describes the fuzziness of the algorithm:
For $m=1$, the algorithm degenerates to the boolean k\nobreakdash-means clustering algorithm, and for the limit $m\to\infty$, all membership values become $u_{ij}=1/C$, yielding a completely fuzzy clustering.
In the range of $m=1.1$ to $m=2.0$ the observed dependence on $m$ of the clustering results is very weak and we use $m=1.2$, which yielded the best results, in the following discussion.
For a more detailed discussion of the $m$-dependence we refer to section 6.1 of \cite{Savran2010}.

Employing Lagrange multipliers, one can derive the necessary conditions for a minimum of $J_m$, \emph{viz.}
\begin{align}
 \vect{c}_j &= \frac{\sum_{i=1}^{N} u_{ij} \vect{x}_i}{\sum_{i=1}^{N} u_{ij}} \\
 u_{ij} &= \Biggl(\sum_{k=1}^{C} \biggl(\frac{\|\vect{x}_i-\vect{c}_j\|}{\|\vect{x}_i-\vect{c}_k\|}\biggr)^{2/(m-1)}\Biggr)^{-1}\; ,
\end{align}
and find a local minimum by iterating these equations starting from randomly chosen $u_{ij}$ obeying the constraints in \eqref{eq:fcm-opt}.
The random choice of the $u_{ij}$ ensures an unbiased application of the clustering procedure.

In the following, we extensively use the isomorphism between a sampled detector signal $s(k)$, where $1\le k\le L$ denotes the sample index, and a vector $\vect{s}$ represented as an $L$-tuple $\vect{s}=(s(k))_{k\in[1,L]}$ with respect to an arbitrarily chosen orthonormal Cartesian basis of $\mathbb{R}^L$.
We nevertheless keep the notation $s(k)$ to emphasize that these signals are functions of (discretized) time.

\section{Experimental setup}
\label{sec:experiment}
The data used to demonstrate the method proposed in this work was taken during a beam time at the Mayer-Leibnitz laboratory (MLL) in Garching near Munich.
The MLL tandem proton accelerator (see e.g.\ \cite{Assmann1974,Laird2013,Wheldon2012}) provided a proton beam with energy of $24\MeV$.
The protons impinged on a $70\,\text{\textmu m}$ $\text{C}\text{D}_2$ foil target.
The elastically and inelastically scattered protons, as well as photons stemming from nuclear reactions inside the foil, were detected by a 8x4 array of CsI(Tl) detectors \cite{CalifaTDR,Alvarez2008,DiJulio2009} at an angle of $37^\circ$ with respect to the beam axis.
The average count rate of a single crystal was of the order of $200\Hz$ resulting in a very small amount of pile-up signals (see also below).
The crystals were read out using Hamamatsu S8664-1010 APDs\footnote{avalanche photo diodes}.
A Mesytec MPRB-16 shaping pre-amplifier, which also provided the bias voltage for the APDs, was used to obtain the detector signal.
This signal was sampled by a $60\MHz$ 12-bit analog-to-digital converter on a FEBEX2 FPGA\footnote{field-programmable gate array} board.
The Multi Branch System data acquisition software \cite{Essel2003} recorded per-event signal traces, each $2000$ samples in length, and additional data generated by the FPGA board for offline analysis.

\section{Signal preparation}
To ensure a stable clustering process, the signal set is prepared by performing baseline subtraction, noise reduction, time alignment and removal of distorted signals.
These steps are described in the following, the procedure is the same as in the application of the method to liquid scintillators \cite{Savran2010}.

First, the signal baseline voltage is estimated by taking an average over the first 100 samples of the trace, where no signal is present, and then subtracted from the signal.
Next, the signals are filtered twice using a FIR\footnote{finite impulse response} low-pass filter with a cutoff frequency of $1.8\,\text{MHz}$.
This is done to reduce noise superimposed on the signal which could impede the clustering.
The cutoff frequency of the low-pass filter is large with respect to the characteristic time scale of the signal: the shortest rise times are larger than $1\,\text{\textmu s}$.
Thus, distortions introduced by this step are minimal.

For the clustering to work, the signals have to be aligned in time.
This alignment is performed using a digital CFD\footnote{constant fraction discriminator}:
First, a search for the signal maximum $r_\text{max} = r(k_\text{max})$ is performed, with $k_\text{max}$ denoting the sample number where the maximum is assumed.
Starting at $k_\text{max}$ and going to earlier times (smaller $k$), the index $k_\text{cfd}$ of the first sample with value less than $0.2\, r_\text{max}$ is determined.
In case of signal pileup, the location of the global signal maximum $k_\text{max}$ is most likely the maximum of the second (piled-up) pulse and the search for the constant fraction point yields a sample number that is far away from the signal maximum.
The constant fraction point, however, will in most cases reside on the rising edge of the first pulse.
Thus, the signal maximum is determined again in the region $[k_\text{cfd}-50,k_\text{cfd}+450]$ and a new constant fraction search is performed.
If the difference between the old and new constant fraction points is larger than $50$ samples, the new point is taken as $k_\text{cfd}$ and the process is repeated.
That way, only the first of the overlapping pulses is selected.
Note that a sub-sample alignment as performed in \cite{Savran2010} is not necessary because the sampling period is small compared to the rise times of the signals.

\label{sec:preparation}
\begin{figure}
 \centering
 \includegraphics{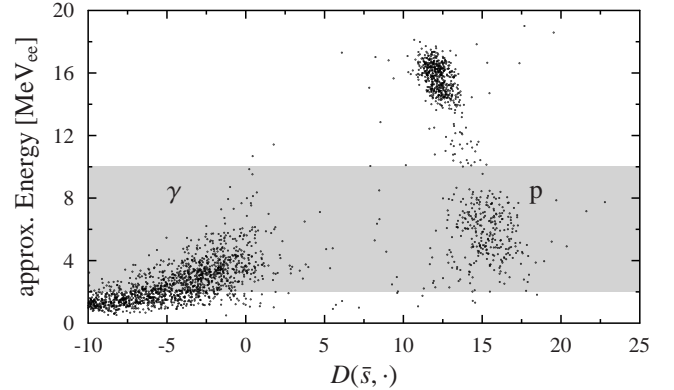}
 \caption{
 Energy deposited in the detector plotted against the distance $D(\bar s,\cdot)$ to the mean signal shape of the full set, given by the $D$ distance measure (see \cref{sec:pid:altern} for details).
 The measure separates both particle species well.
 In the region above $8\MeVee$ only proton signals are present, whereas photon signals dominate in the region below $4\MeVee$.
 The (shaded) region between $2\MeVee$ and $10\MeVee$ was selected because it contains a sufficient number of signals of each signal class.
 }
 \label{fig:eoverdmean}
\end{figure}

The energy of the event is then estimated by taking the average of the five samples at $k_\text{max}-2,\dotsc,k_\text{max}+2$.
Because the clustering results deteriorate when the ratio between photon- and proton-induced signals is too far from unity \cite{Savran2010}, the energy range between $2\MeVee$ and $10\MeVee$, which contains a comparable number of signals from both species, is selected (see \cref{fig:eoverdmean}).
All signals whose energies lie outside this region are removed from the set.
Since only the proton pulse shape shows an energy dependence, it would be possible to extend the clustering to higher energies by first applying the fuzzy c\nobreakdash-means to the selected region which contains both types of signals and then running the algorithm in the region with $E>10\MeVee$ only for the proton cluster while keeping the photon cluster centroid fixed.

The signals are then cropped so that only 50 samples before and 450 samples after $k_\text{cfd}$ are retained.
This ensures that the constant fraction point is the 51\textsuperscript{st} sample in all cropped signals $s_i(k)$.
The cropped signals are normalized such that the signal energy is equal to unity.

Because a contamination of the signal set with distorted signals (e.g.\ due to pile-up or clipping) may lead to inferior clustering results, these signals have to be eliminated (see \cite{Savran2010} for details).
Therefore, a mean signal shape
\begin{equation}
 \bar s(k) = \frac{1}{N} \sum_{i=1}^{N} s_i(k)\label{eq:meansignal}
\end{equation}
is calculated, where $N$ is the number of signals still in the set.
All signals with a euclidean distance larger than $2.7$ to this mean shape are considered distorted and eliminated from the set.
This treatment was sufficient for the signal set at hand, removing only about $0.5\%$ of the signals.
For sets with a larger number of distorted signals a more sophisticated method, such as hierarchical clustering, may be necessary.

After performing the correction of the CFD timings, pileup did not play a significant role for the present signal set.
Any signal that contains a piled-up pulse starting inside the cropped region is considered distorted and discarded by the last step.
Piled-up pulses starting after the cropped region are ignored.

\section{Particle identification}
\label{sec:pid}
In the following sections, two methods for achieving particle identification are presented, the first of which is the well-known integration method.
The second method uses fuzzy clustering to determine a principal pulse shape for each particle species and may be combined with various discrimination functions that make use of the principal pulse shape.
Two discrimination functions are presented and their performance is compared to the performance of the integration method.

The comparison is done using the signals of two neighboring detector crystals near the center of the 8x4 array.
We show a PID histogram for each of the investigated methods and each detector and, in order to provide a benchmark, we give a Figure-of-Merit (FOM) in the respective figures.
The FOM is defined as
\begin{equation}\label{eq:pid:fom}
 M=\frac{|\mu_1 - \mu_2|}{\sigma_1+\sigma_2}\;,
\end{equation}
where $\mu_1,\mu_2$ and $\sigma_1,\sigma_2$ are the means and standard deviations of gaussian functions fitted to the two peaks of each histogram.
The FOM is thus a mesure of how well the respective method separates the two particle species for the data set at hand.

\subsection{Integration method}
\label{sec:pid:integr}
\begin{figure}
 \centering
 \includegraphics{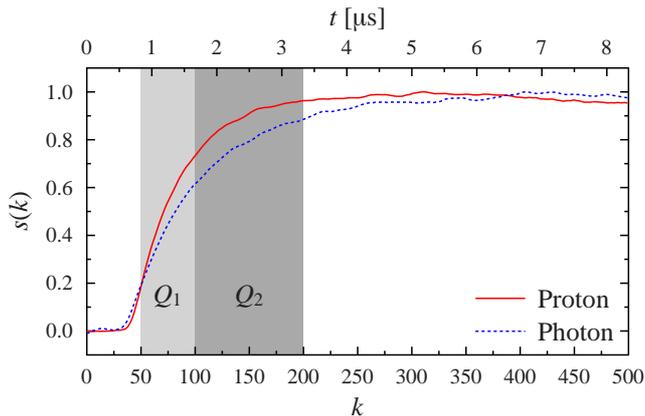}
 \caption{(color online)
 Typical normalized single signal shapes for protons and photons of a CsI(Tl) detector with APD readout passed through a shaping amplifier.
 The shaded areas mark the integration limits used by the integration method.
 }
 \label{fig:integrlim}
\end{figure}
\begin{figure}
 \centering
 \includegraphics{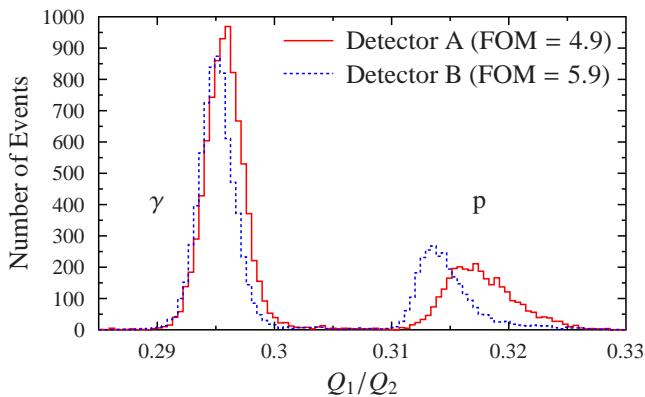}
 \caption{(color online)
 $Q_1/Q_2$ histograms for two different detectors.
 The peaks are well-separated, but their positions in both histograms differ slightly due to different characteristics of the pulse shapes for the two detectors.
 }
 \label{fig:qqspectrum}
\end{figure}
\begin{figure}
 \centering
 \includegraphics{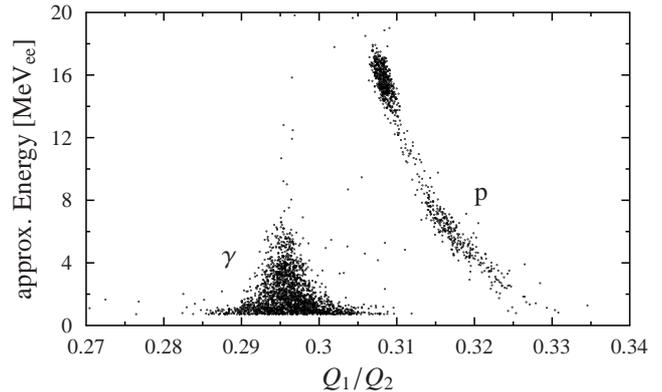}
 \caption{
 Energy deposited in the detector plotted against the $Q_1/Q_2$ ratio used by the integration method to discriminate particle species.
 The proton signal shapes show an energy dependence in the $Q_1/Q_2$ ratio while the photon shapes are largely independent of energy.
 }
 \label{fig:eoverqq}
\end{figure}
Because different particle species induce distinct processes in the scintillator, the signal shapes will differ and thus the integral over a part of the signal will give different values depending on the particle species that induced it.
The ratio of two integrals
\begin{equation}
 \frac{Q_1}{Q_2}=\frac{\sum_{k=a_1}^{b_1} s(k)}{\sum_{k=a_2}^{b_2} s(k)}
\end{equation}
over two different regions $[a_1,b_1]$ and $[a_2,b_2]$ is taken to make the value independent of the total integral and thus usable for particle identification.
A histogram of the $Q_1/Q_2$ ratios for the set of signals is then created. Each particle species in the signal set produces a peak in the histogram at a characteristic $Q_1/Q_2$ ratio and cuts can be applied to perform particle identification.
In general, the integration limits have to be adjusted to the characteristics of the detector signals to get good identification results.

The integration limits for the signal set at hand are illustrated in \cref{fig:integrlim}.
The $Q_1/Q_2$ histogram calculated using these limits is depicted in \cref{fig:qqspectrum} for two different detectors.
The peaks in the two histograms have different positions for the indiviual detectors (due to slightly different characteristics of the pulse shapes) and these positions cannot be known beforehand.
They have to be determined by using a fitting procedure or by manual intervention.
The FOM values for the two detectors are also given in \cref{fig:qqspectrum}.
Due to the wide proton peak in the histogram for detector A the value is significantly smaller.

\Cref{fig:eoverqq} shows the signal energy as a function of the $Q_1/Q_2$ ratio for the signals in the present set.
The photon signal shape is largely independent of energy.
However, the shape of the proton signals shows an energy dependence, causing an energy-dependent change in the $Q_1/Q_2$ ratio.

\subsection{Fuzzy clustering}
\label{sec:pid:fuzzy}

\begin{figure}
 \centering
 \includegraphics{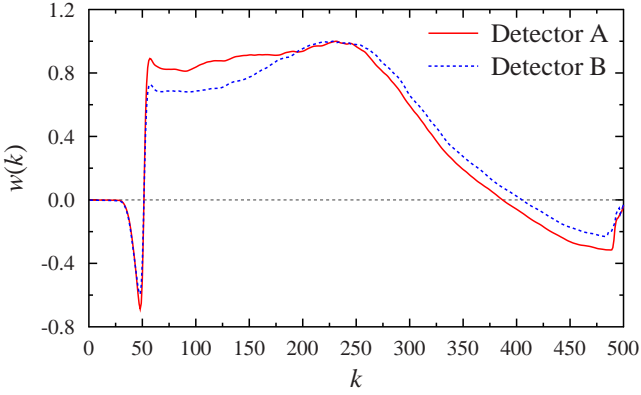}
 \caption{Weighting functions obtained from \eqref{eq:pid:fuzzy:weight} for two detectors, used for the calculation of the amplitude histograms depicted in \cref{fig:dvspectrum}.}
 \label{fig:weightingfunctions}
\end{figure}
\begin{figure}
 \centering
 \includegraphics{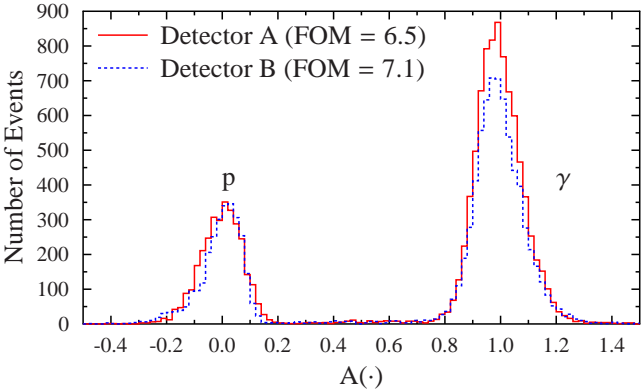}
 \caption{(color online)
 Histograms of discrimination values for two detectors. The histograms are calculated from the same signal sets as the $Q_1/Q_2$ ratios from \cref{fig:qqspectrum}.
 The proton and photon peaks are situated around a discrimination value of $0$ and $1$ respectively and the proton and photon signals form completely disjunct distributions.
 }
 \label{fig:dvspectrum}
\end{figure}

The fuzzy clustering method achieves particle identification by comparing detector signals to prototype signals for each particle species.
These prototype signals do not have to be constructed manually, but can be derived from a mixed set containing a large number of signals originating from the different species.
This makes the fuzzy clustering method a two-step procedure, with the derivation of the prototype signals as first and the assignment of individual signals to particle species as second step.

The first step is performed by applying the fuzzy c-means clustering algorithm to the signal set.
The clustering yields a prototype signal $c_j(k)$, given by the determined cluster centroids, for each particle species and a matrix of membership values $u_{ij}$ denoting the degree of membership in each cluster for each signal.
The intra-cluster variance per sample $k$
\begin{equation}
 \Delta^2_j(k) = \frac{1}{U_j}\sum_{i=1}^{N} u_{ij} (s_i(k) - c_j(k))^2 \;,
\end{equation}
with the total cluster size $U_j = \sum_i u_{ij}$, and other statistical quantities can easily be calculated from these results.

The membership values can be used themselves to perform particle identification.
However, they tend to values very close to unity or zero when the fuzziness parameter $m$ is small and are thus hardly comparable to the results of other particle identification methods.

It is thus desirable to derive a different means for achieving particle identification from the results of the clustering process.
The information obtained from the clustering method facilitate the use of an improved integration method.
Instead of setting the integration regions manually, the prototype signals can be used to determine a set of weighting functions 
\begin{equation}\label{eq:pid:fuzzy:weight}
w_{ij}(k) = \frac{c_i(k) - c_j(k)}{\Delta^2_i(k) + \Delta^2_j(k)}\;.
\end{equation}
These weights are largest where the prototype signals $c_i(k)$ of the clusters under consideration differ the most with respect to their variances $\Delta^2_i(k)$.
As there is considerable scatter inside each cluster, the widths of both signal distributions have to be taken into account.
This is achieved by relating the difference between the prototype signals to the sum of the intra-cluster variances of both clusters.
The weights obtained for the present data set are shown in \cref{fig:weightingfunctions}.

Using the weights $w_{ij}(k)$, a discrimination value for each pair of particle species
\begin{equation}
 \tilde A_{ij} [s(k)] = \sum_{k=0}^L w_{ij}(k) s(k)\;,
\end{equation}
with $L$ being the length of the signal in samples, can be assigned to each signal from the set.
This step can be expressed as a scalar product of two vectors and is suitable for implementation on fast digital electronics.

The discrimination value $\tilde A_{ij}$ for the two prototype signals can also be calculated.
This is advantageous, because the discrimination values can be normalized so that a value of zero and unity is assigned to both prototype signals respectively:
\begin{equation}
 A_{ij} [s(k)] = \frac{\tilde A_{ij}[s(k)] - \tilde A_{ij}[c_j(k)]}{\tilde A_{ij}[c_i(k)] - \tilde A_{ij}[c_j(k)]}\;.
\end{equation}
With the help of this information, the peak positions in the discrimination value histogram corresponding to the two clusters are $A_{ij}=0$ and $A_{ij}=1$ by definition.
The widths of the peaks can be estimated using the intra-cluster variance of the discrimination values
\begin{equation}
 \Delta A^2_{ij}(l) = \frac{1}{U_l}\sum_{n=1}^{N} u_{nl} (A_{ij}[s_n(k)] - A_{ij}[c_l(k)])^2\;,
\end{equation}
which describes the variance of the peak produced by cluster $l$ in the histogram of discrimination values computed with respect to clusters $i,j$.
These initial guesses can be used to perform an automated least-squares fit to the histogram, which should converge rapidly due to the accuracy of the initial guesses.
Then, cuts can be automatically derived from the fitted distributions.

\Cref{fig:dvspectrum} shows the histograms of discrimination values for the same signal sets as in \cref{fig:qqspectrum}.
The peaks corresponding to protons and photons are situated near zero and one, as expected.
The peak positions are also identical for both detectors, even though the pulse shapes for the two detectors differ slightly, leading to different distributions in the $Q_1/Q_2$ histogram (see \cref{fig:qqspectrum}).
Compared to the integration method also the FOM is slightly improved (compare the given values in the respective figures).
This shows that our approach of defining the weighting function without prior knowledge of the properties of the signal shapes gives reasonable results.

The prescription for obtaining the discrimination amplitudes can be interpreted as a special case of Fisher's linear discriminant for vanishing covariance between the signal samples described in \cite{Duda2000}.
Although that assumption is not strictly true due to the initial low-pass filtering of the signals, which induces significant correlations between neighboring samples, the results provide an excellent means for discriminating between the particle species in an unsupervised way.

A similar idea has been pursued earlier for photon-alpha discrimination in \cite{Skulski2001}: their method D consists of a weighted integration over the signal to obtain a value that can be used for discrimination.
The difference to the procedure proposed in this work is how the weights are obtained: by averaging over two separate sets of known photon and alpha signals, the authors obtain two prototype signals.
The difference between these two signals is then used to construct the weight by fitting a polynomial to the rising edge of the difference and attaching an exponential tail to it.
The tail decay constant is then optimized for maximum cluster separation.

Using fuzzy clustering to obtain the prototype signals renders the need for a separate calibration set for each particle species unnecessary.
Fisher's linear discriminant ensures optimal discrimination independent of the specific form of the difference signal.

\subsection{Distance approach}
\label{sec:pid:altern}
\begin{figure}
 \centering
 \includegraphics{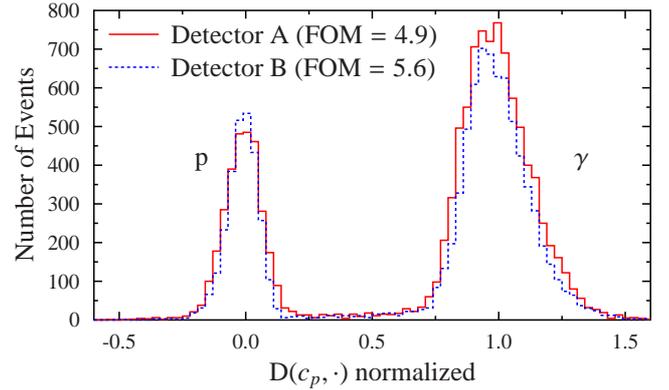}
 \caption{(color online)
   The scaled distance $D$ to the proton prototype signal for the same data as in \cref{fig:qqspectrum} and \cref{fig:dvspectrum}.
   Using the value of $D$ for the photon prototype signal the distance can be scaled to result in $0$ and $1$ respectively.
 }
 \label{fig:dpscaled}
\end{figure}

Instead of performing a weighted integration over the signal, a notion of the distance between two signals can be used to perform particle identification.
Interpreting the signals as vectors in a high-dimensional vector space, the most straight-forward way of defining a metric is using the euclidean distance between the two signal vectors
\begin{equation}
 d(s_1,s_2) = \sqrt{(s_1 - s_2)^T (s_1 - s_2)}\;.
\end{equation}
This naive approach, however, has some drawbacks, because it does not take the properties of real signals into account.
The signals contain noise which is accumulated when calculating the euclidean distance and may significantly increase the distance between two otherwise very similar signals, producing false results.

A way to circumvent this has been employed in \cite{Savran2010}: a distance measure
\begin{equation}
 \label{eq:alt:Dmeasure}
 D(s_1,s_2) = \sum_k (s_1(k) - s_2(k))
\end{equation}
is defined, which allows for noise contributions to cancel between different samples.
This distance measure can then be used to calculate a discrimination amplitude based on the distance to one of the cluster prototype signals.
Again, the amplitudes of the other prototype signals can be calculated and used to scale the discrimination amplitude so that the prototypes are at zero and unity, respectively.
\Cref{fig:dpscaled} shows a histogram of the distances to the proton cluster centroid $D(c_p,\cdot)$, scaled to obtain a plot similar to the discrimination value histogram shown in \cref{fig:dvspectrum}.
Compared to the method using the weighting function the determined values of the FOM for the distance approach are slightly worse and comparable to the integration method.
This is probably not surprising since the weighting function explicitly concentrates on the region of maximum distance between the clusters, taking statistical fluctuations into account.

We want to mention in this context one peculiarity of the $D$ measure, which is that the distances between three points are tightly related:
\begin{align}
 D(s_1,s_2) &= \sum_k (s_1(k) - s_2(k)) \notag\\
 &= \sum_k (s_1(k) - s_3(k) + s_3(k) - s_2(k)) \notag\\
 &= D(s_1,s_3) + D(s_3,s_2)\;.
\end{align}
A change of the reference signal for the calculation of the discrimination amplitude thus results only in a (irrelevant) constant shift of the amplitudes.
For the two-species case this can be exploited by taking the overall mean signal \eqref{eq:meansignal} as a reference, effectively removing the need for performing the clustering.
In addition to that, no care has to be taken that the ratio of the signals induced by each of the particle species under consideration is not too far from unity.
The drawback is that the peak positions and widths are not known beforehand anymore.
We used this approach in \cref{fig:eoverdmean} in order to select the energy region over which the clustering is performed.

\section{Conclusions and outlook}
We propose a robust and reliable method to perform PID calibration of a detector in an unsupervised manner.
It is applied to a data set of mixed photon- and proton-induced signals taken with CsI(Tl) detectors at moderate count rates and similar population of proton and photon signals.
This method uses fuzzy clustering algorithms to obtain prototype pulse shapes for each particle species in a mixed set of signals, which are used to derive a discrimination amplitude for each signal.
The calculation of the discrimination values is not computationally expensive, i.e.\ once the weighting functions have been determined for a given detector they can be used to calculate the discrimination amplitudes for further data online using fast digital electronics.
By recording detector signals for a small percentage of events and regularly performing clustering on this set, a change of the prototype pulse shapes (e.g.\ caused by a change of conditions) can be detected and acted upon by updating the discrimination amplitude.

The amplitudes derived this way are used to assign signals to particle species using discrimination regions that are---in contrast to other methods---detector-independent.
Furthermore, the results show that using the derived weighting functions the Figure-of-Merit, a measure of the separation between particle species, is comparable to the integration method.

In an earlier work \cite{Savran2010} we successfully applied the method to $\gamma$-neutron discrimination in liquid scintillator detectors.
In this work, we show that the same prescription can be used to achieve particle identification for $\gamma$-proton discrimination in CsI(Tl) scintillators.
This shows that the proposed method is indeed a generic one, which can be applied to many different detector types.
The method is in principle also applicable for more than two particle species.
This case should be investigated in upcoming works.

\section*{Acknowledgements}
This work was supported by the Alliance Programme of the Helmholtz Association (HA216/EMMI), BMBF (06DA9040I and 05P12RDFN8) and HIC for FAIR.




\section*{References}
\bibliographystyle{apsrev}
\bibliography{fcm-paper.bib}







\end{document}